\documentclass[aps,twocolumn,superscriptaddress]{revtex4}
\usepackage{float}
\usepackage{graphicx}
\usepackage{dcolumn}
\usepackage{bm}
\usepackage{color}
\usepackage{amsmath}
\usepackage{amssymb}
\usepackage{hhline}
\usepackage[normalem]{ulem}
\usepackage[dvipsnames]{xcolor}

\begin{document}
\title{Relative importance of nonlinear electron-phonon coupling and vertex 
	corrections in the Holstein model}
\author{Philip M. Dee}
\affiliation{Department of Physics and Astronomy, The University of Tennessee, 
Knoxville, Tennessee 37996, USA}
\author{Jennifer Coulter}
\affiliation{John A. Paulson School of Engineering and Applied Sciences, Harvard 
University, Cambridge, MA, USA}
\author{Kevin Kleiner}
\affiliation{Department of Physics and Astronomy, The University of Tennessee, 
Knoxville, Tennessee 37996, USA}
\author{Steven Johnston}
\email{sjohn145@utk.com}
\affiliation{Department of Physics and Astronomy, The University of Tennessee, 
Knoxville, Tennessee 37996, USA}
\affiliation{Joint Institute for Advanced Materials at the University of 
Tennessee, Knoxville, Tennessee 37996, USA}
\date{\today}

\begin{abstract}
%------------------------------------------------------------------------------%
\section*{Abstract}
Determining the range of validity of Migdal's approximation for electron-phonon 
({\it e}-ph) coupled systems is a long-standing problem. 
Many attempts to answer this question employ the Holstein Hamiltonian, where the 
electron density couples linearly to local lattice displacements. 
When these displacements are large, however, nonlinear corrections to the 
interaction must also be included, which can significantly alter the physical 
picture obtained from this model. 
Using determinant quantum Monte Carlo and the self-consistent Migdal 
approximation, we compared superconducting and charge-density-wave correlations 
in the Holstein model with and without second-order nonlinear interactions. 
We find a disagreement between the two cases, even for relatively small values 
of the {\it e}-ph coupling strength, and, importantly, that this can occur in the 
same parameter regions where Migdal's approximation holds. 
Our results demonstrate that questions regarding the validity of Migdal's 
approximation go hand in hand with questions of the validity of a linear 
{\it e}-ph interaction. 
\end{abstract}

%------------------------------------------------------------------------------%
%---------------------------- Introduction-------------------------------------%
%------------------------------------------------------------------------------%
\maketitle
\section*{Introduction}
Our modern understanding of phonon-mediated superconductors is largely based on 
results from ab initio approaches~\cite{Giustino2017} coupled with 
Migdal's approximation~\cite{Migdal1958,Eliashberg1960,Eliashberg1961}. 
Migdal's approximation~\cite{Migdal1958} neglects corrections to the 
electron-phonon ({\it e}-ph) interaction vertex, which scale as 
$O(\lambda\frac{ \hbar\Omega }{ E_{\text{F}} })$, where $ \lambda $ is a 
dimensionless measure of the {\it e}-ph coupling strength, $ \hbar\Omega $ is the 
typical phonon energy, and $E_{\text{F}}$ is the Fermi energy. 
Physically, this approximation neglects processes leading to polaron formation, 
and determining precisely when these processes become important and their impact 
on transport properties is a long-standing problem~\cite{Freericks1997,
Alexandrov2001,Hague2003,Bauer2011,Esterlis2018,Liu2019,Schrodi2019,
Gastiasoro2019}. 

%------------------------------------------------------------------------------%
Many attempts to address this question have utilized nonpertubative simulations 
of simplified effective models like the Holstein~\cite{Holstein1959} or 
Fr\"{o}hlich~\cite{Frohlich1954} Hamiltonians, where the electron density 
couples linearly with phonon fields. 
For example, owing to it's relative simplicity, the Holstein model and its 
extensions have been studied extensively using quantum Monte Carlo 
(QMC)~\cite{Scalettar1989,Marsiglio1990,Levine1990,Levine1991,Noack1991,Vekic1992,Vekic1993,Niyaz1993,Freericks1995PRL,Freericks1997,FreericksPRL1997,Zheng1997,Goodvin2006,Chen2018,Li2019,Hohenadler2019}, and 
serves as a prototype for studying different polaronic regimes. 
Recently, it was shown that even if $\frac{ \hbar\Omega }{ E_{\text{F}} } < 1$, 
one can find instances where vertex corrections (i.e. polaron formation) become 
important for $\lambda\approx 0.4$ -- $0.5$~\cite{Esterlis2018,Bauer2011}. 

%------------------------------------------------------------------------------%
It is generally understood that small (large) polarons form when the polaron 
binding energy is larger (smaller) than the hopping energy of the 
carriers~\cite{Devreese1996}. 
However, small polarons are often accompanied by sizable lattice distortions and 
a tendency toward localization and charge order.  
This observation has motivated some work to include higher-order nonlinear 
{\it e}-ph coupling terms to study changes in polaron formation~\cite{Adolphs2013} 
and on charge-density-wave (CDW) and superconducting (SC) pairing 
correlations~\cite{Li2015a,Li2015b}. 
These studies found that small positive (negative) nonlinear terms decrease 
(increase) the effective mass of the carriers and contracts (enlarges) the local 
lattice distortions surrounding the carriers~\cite{Adolphs2013}. 
Furthermore, mean-field treatments aiming to recover a linear model via 
effective model parameters fail to capture the quantitative nature of the true 
nonlinear model~\cite{Adolphs2013,Li2015a,Li2015b}, indicating that 
nonlinearities cannot be renormalized out of the problem. 
For example, one can tune the parameters of an effective linear model to 
capture either the electronic or phononic properties of the nonlinear model but 
not both simultaneously~\cite{Li2015b}. 
This failure is important to note in the context of polaron formation, where the 
electrons and phonons become highly intertwined. 
To capture this physics accurately, an effective model must describe both 
degrees of freedom on an equal footing, and an effective linear description of 
a nonlinear {\it e}-ph model will not do this.

%------------------------------------------------------------------------------%
These results raise an important question about the priority of investigations 
into the validity of the aforementioned approximations. 
Are there scenarios where the breakdown of the linear approximation supersedes 
the breakdown of Migdal's approximation? 
In this work, we show that this is indeed the case. 
Specifically, by comparing QMC simulations of the (non)linear Holstein 
model with results obtained with the Migdal approximation's, we show that 
nonlinear corrections can be more important than vertex corrections, and that 
this can occur even when Migdal's approximation appears to be valid. 
Our results have consequences for any conclusions drawn about the validity of 
Migdal's approximation from model Hamiltonians and highlight a critical need to 
move beyond such models for a complete understanding of strong 
{\it e}-ph interactions. 

%------------------------------------------------------------------------------%
%------------------------------- Results --------------------------------------%
%------------------------------------------------------------------------------% 
\section*{Results}
\noindent{\bf Model}. We study an extension of the Holstein model that includes 
nonlinear {\it e}-ph interaction terms and defined on a two-dimensional (2D) square 
lattice. 
The Hamiltonian is 
$ \hat{H} = \hat{H}_{\text{el}} + \hat{H}_{\text{lat}} + \hat{H}_{\text{int}} $, 
where 
\begin{equation}\label{eq:H_el}
	\hat{H}_{\text{el}} = 
	-t \sum_{\langle i, j \rangle, \sigma}\hat{c}^{\dagger}_{i,\sigma}\hat{c}^{{\phantom{\dagger}}}_{j,\sigma}
	-\mu\sum_{i}\hat{n}_{i}	
\end{equation}
and
\begin{equation}\label{eq:H_lat}
	\hat{H}_{\text{lat}} = \sum_{i}\left(\frac{\hat{P}_i^2}{2M} 
	+ \frac{M\Omega^2 \hat{X}_i^2}{2}\right)
	= \sum_{i}\hbar\Omega\left(\hat{b}^{\dagger}_{i}\hat{b}^{{\phantom{\dagger}}}_{i} + \frac{1}{2}\right)
\end{equation}
describe the noninteracting electronic and phononic parts, respectively, and
\begin{equation}\label{eq:H_int}
	\hat{H}_{\text{int}} = 
	\sum_{i, k} \alpha^{\phantom\dagger}_{k} \hat{n}^{\phantom\dagger}_{i} 
	\hat{X}_{i}^{k} 
	= \sum_{i, k }g^{\phantom\dagger}_{k}\hat{n}^{\phantom\dagger}_{i}
	\left(\hat{b}^{\dagger}_{i} + \hat{b}^{{\phantom{\dagger}}}_{i}\right)^{k}
\end{equation} 
describes the {\it e}-ph interaction to $k$\textsuperscript{th} order in the atomic 
displacement.
Here, $\hat{c}^{\dagger}_{i,\sigma}$ ($\hat{c}^{{\phantom{\dagger}}}_{i,\sigma}$) 
create (annihilate) spin $\sigma$ $(= \uparrow,\downarrow)$ electrons on site 
$ i $, 
$ \hat{n}_{i} =\sum_{\sigma}\hat{c}^{\dagger}_{i,\sigma}\hat{c}^{{\phantom{\dagger}}}_{i,\sigma} $ is the number operator, $\mu$ is the chemical potential, $t$ is the nearest-neighbor 
hopping integral, and $\langle i,j \rangle$ restricts the summation to nearest 
neighbors only.
Each ion has a mass $ M $ with position and momentum operators denoted by 
$\hat{X}_i$ and $\hat{P}_i$, respectively.
Quantizing the lattice vibrations leads to Einstein phonons created 
(annihilated) by the operator 
$\hat{b}^{\dagger}_{i}$ ($\hat{b}^{{\phantom{\dagger}}}_{i}$) with associated phonon 
frequency $ \Omega $. 
Lastly, $\alpha_k$ and $g_k$ are the {\it e}-ph interaction strengths in the two 
representations, which are related by 
$ g_{k} = \alpha_{k}\left(\frac{\hbar}{2M\Omega}\right)^{\frac{k}{2}} $. 

%------------------------------------------------------------------------------%
Following previous works on the nonlinear Holstein model~\cite{Li2015a,Li2015b}, 
we truncate the series in $\hat{H}_{\text{int}}$ to second order and introduce 
the ratio $ \xi = g_{2}/g_{1} $ to quantify the relative size of the two 
{\it e}-ph couplings. 
(The standard Holstein model is recovered by setting $ g_{2} = 0 $.)
This simplification is sufficient to assess the relative importance of 
nonlinear interactions relative to Migdal's approximation. 
Additional orders up to $k=4$ have been studied in the single carrier 
limit~\cite{Adolphs2013}, where they produce the same qualitative picture.
%

%------------------------------------------------------------------------------%
To facilitate comparison with previous work, we set $k_B = \hbar = t = M = 1$. 
The scale of atomic displacements is set by the oscillation amplitude of the free harmonic oscillator  
$A = \sqrt{1/2\Omega}$ ($\sqrt{\hbar/2M\Omega}$ with the physical units 
restored).
When reporting expectation values of $X$ and its fluctuations, we 
explicitly divide by $A$ in model units, thereby making the results dimensionless.
To get an idea for what these values mean in reality, one can simply 
multiply the results by $A$ in physical units.
Later, we will consider FeSe to estimate the strength of the nonlinear 
interactions.
In that case, the prefactor (in physical units) is $A\sim 0.036$~\AA{}, 
which is obtained after adopting a selenium mass $M = 1.31\times 10^{-25}$ kg 
and the experimental phonon energy $\hbar \Omega = 20.8 $ meV of the A$_{1g} $ 
mode (see Supplementary Note 1 for more details). 
Alternatively, we obtain a comparable scale of $A \sim 0.051$~\AA{} 
for the transition metal oxides, where $\hbar\Omega = 50$ meV and 
$M=2.66\times 10^{-26}$ kg are typical for the optical oxygen phonons.
Finally, we adopt the standard definition for the dimensionless 
\emph{linear} {\it e}-ph interaction strength $ \lambda = 2g_{1}^{2}/W\Omega $,
where $ W=8t $ is the bandwidth. 

In what follows, we compare results obtained using determinant quantum Monte 
Carlo (DQMC)~\cite{WhitePRB1989} and the self-consistent Migdal approximation 
(SCMA)~\cite{Marsiglio1990} (see Methods and Supplementary Note 2). 
To determine the relative importance of nonlinear {\it e}-ph interactions against 
vertex corrections to Migdal's approximation, we juxtapose results obtained 
using these methods for the linear and nonlinear models.
For example, comparing results obtained from DQMC and the SCMA for the linear 
model reveals the importance of vertex corrections. 
Analogously, comparing DQMC results for the linear and nonlinear models provides 
a measure for the importance of nonlinear interactions while treating the two 
models exactly. 
This methodology will allow us to isolate the source of any observed 
discrepancies. 
Deviations between SCMA and DQMC for the linear model must be due to vertex 
corrections, while disagreement between DQMC results for the linear and 
nonlinear models must arise from the additional quadratic interaction.
%------------------------------------------------------------------------------%

\begin{figure}[t]
	%	\centering
	\includegraphics[width=1.0\columnwidth]{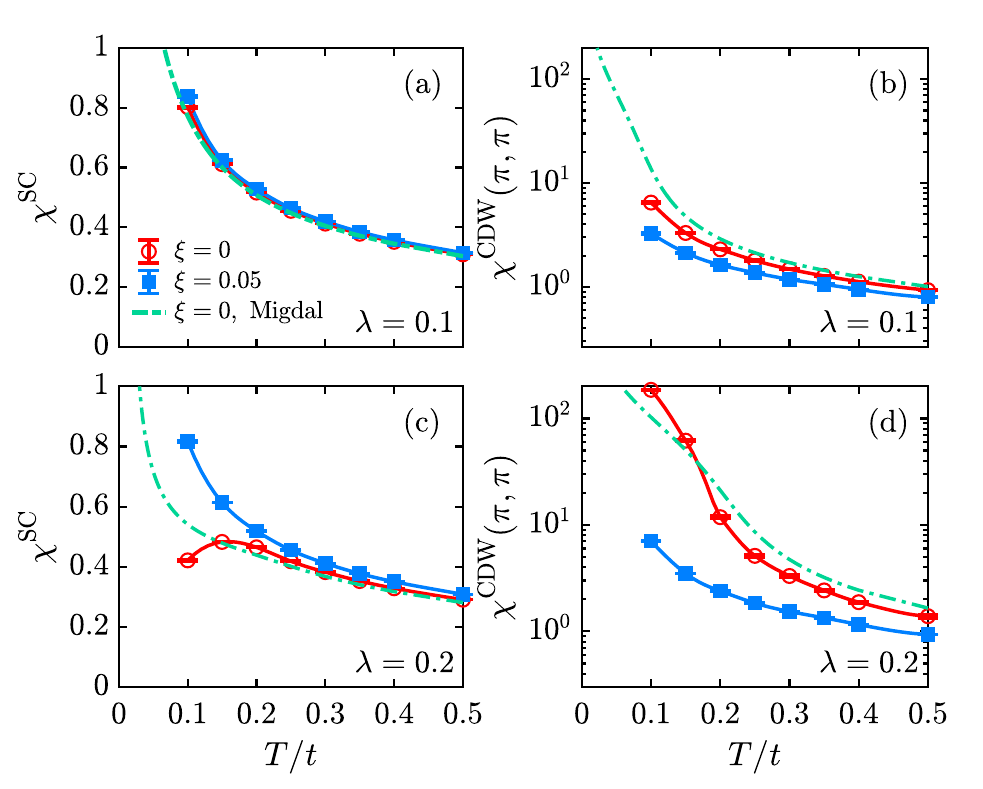}
	\caption{
		\textbf{Comparison of the superconducting (SC) and charge 
		susceptibilities at half-filling.}
		The singlet-pairing ($\chi^{\text{SC}}$) and charge-density-wave   
		($ \chi^{\text{CDW}}(\pi,\pi) $) susceptibilities vs. temperature for 
		dimensionless {\it e}-ph couplings of $ \lambda =0.1 $ (panels 
		\textbf{a}-\textbf{b}) and 0.2 (panels \textbf{c}-\textbf{d}) at 
		half-filling and $ \Omega =0.5t $.
		Results for the model with and without nonlinear corrections are 
		shown using closed and open symbols, respectively.
		Error bars on the DQMC data points are one standard deviation 
		statistical errors estimated using jackknife resampling.
		}
	\label{fig:1}
\end{figure}
%
%------------------------------ half-filling ----------------------------------%
\vskip 0.2cm
\noindent{\bf Comparison of susceptibilities at half-filling}. 
We begin by comparing the susceptibilities for charge-density-wave (CDW)   
$ \chi^{\text{CDW}}(\pi,\pi) $ and pairing $ \chi^{\text{SC}} $ correlations 
for a few illustrative cases (see Methods). 
The first comparison takes place at half-filling 
$ n \equiv \langle\hat{n}_{i}\rangle = 1 $,  
where both CDW correlations and lattice displacements are significant. 
For example, at $\lambda$ = 0.2, $\Omega = 0.5t$, and $T = 0.1t$, we obtain 
$ |\langle X_{i,l} \rangle|/A \sim 1.97 $ and $2.53$ for $\xi=0.05$ and 
$\xi=0$, respectively.
Taking $A\approx 0.036$~\AA{} for FeSe, these values correspond to 
approximately 2.4\% and 3.1\% of the $2.95$~\AA{} Fe-Fe bond  length. 
Similarly, taking $A\approx 0.051$~\AA{} translates to 5.1\% and 6.6\% of the 
typical $1.96$ \AA{} Cu-O bond-distance in a high-$T_{\text{c}}$ superconducting 
cuprate.
These displacements are not negligible (as we will show) when the 
nonlinearities are included, particularly given the weak values of the 
coupling we consider here. 
Later, we also discuss the size of the corresponding vibrational 
fluctuations.

Fig.~\ref{fig:1} presents results for the temperature dependence of 
$ \chi^{\text{CDW}}(\pi,\pi) $ and $ \chi^{\text{SC}} $ using 
$ \Omega = 0.5t $, $ N = 8 \times 8  $, and $ \lambda = $ 0.1 and 0.2. 
This parameter set corresponds to weak coupling and satisfies the adiabatic 
criterion $\frac{\Omega}{E_\mathrm{F}} < 1$, where we expect the SCMA to hold. 
Indeed, when $ \lambda = 0.1 $ (Fig.~\ref{fig:1}a-b), there is fair agreement 
between the DQMC results obtained from both the linear ($ \xi = 0 $) and 
nonlinear ($ \xi = 0.05 $) {\it e}-ph models (symbols with solid curve), as well as 
the SCMA results for the linear model (dash-dot curve). 
When $ \lambda = 0.2 $ (Fig.~\ref{fig:1}c-d), however, we find significant 
disagreement between the results for $ \xi = 0 $ and $ \xi = 0.05 $ in both 
susceptibilities, especially at lower temperatures.
In Fig.~\ref{fig:1}d the DQMC and SCMA results mostly agree for the linear 
Holstein model ($ \xi = 0 $), but a small nonlinear correction of 
$ \xi  = 0.05 $ yields a marked suppression the CDW correlations.
The rapid onset of CDW order in the $ \xi = 0 $ case (Fig.~\ref{fig:1}d) 
coincides with a sharp downturn in $ \chi^{\text{SC}} $  (Fig.~\ref{fig:1}c), 
a feature which isn't captured by the SCMA result.

%------------------------------------------------------------------------------%

\begin{figure}[t]
	\includegraphics[width=1.0\linewidth]{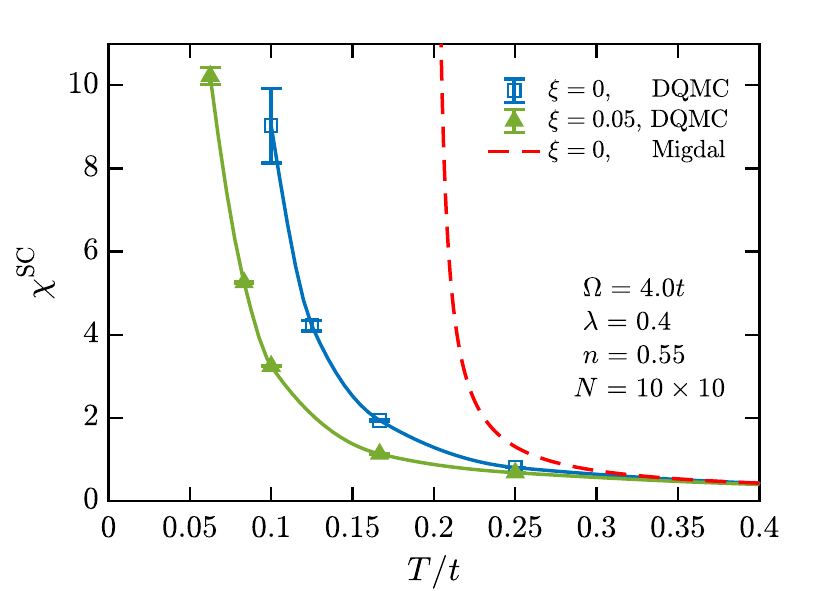}
	\caption{\textbf{Temperature dependence of the superconducting 
	susceptibility 	for a large phonon frequency $\boldsymbol{ \Omega = 4.0t }$.}
		Results are shown for a filling of $ n =0.55 $ and a lattice size 
		$ N = 10\times10 $.
		Both the linear ($ \xi = 0 $, blue squares) and nonlinear 
		($ \xi = 0.05 $, green triangles) Holstein model results from 
		determinant quantum Monte Carlo (DQMC) show a rapid growth of pairing 
		correlations with decreasing temperature, but approach different 
		asymptotes.  
		The self-consistent Migdal approximation (SCMA) results (red dashed 
		line), shown here for reference, yield a large and inaccurate estimate 
		for the superconducting critical temperature due to the invalidity of 
		Migdal's approximation.  
		The lines connecting DQMC data are spline-interpolated and used only to 
		guide the eye.
	    Error bars on the DQMC data points are one standard deviation 
		statistical errors estimated using jackknife resampling.
	}
	\label{fig:2}
\end{figure} 

%------------------------------------------------------------------------------%
The suppression of CDW correlations (Fig.~\ref{fig:1}d) in the presence of 
nonlinear {\it e}-ph coupling demonstrates the importance of higher-order 
interactions over vertex corrections in this case.
As we show later, the need for nonlinear {\it e}-ph coupling is greatest near 
half-filling, where the CDW correlations are strongest. 
Of course, the downturn of the pairing susceptibility obtained from DQMC 
($ \xi = 0 $) at lower temperatures (Fig.~\ref{fig:1}c) appears to indicate 
that vertex corrections are also important for capturing the low temperature 
behavior of $\chi^{\text{SC}}$ at $\lambda \sim 0.2$.
This value of $\lambda$ is smaller than the breakdown values reported 
in Esterlis \textit{et al}.~\cite{Esterlis2018}, however, our models differ 
slightly.
For one, they suppress the effects of Fermi-surface nesting by situating the electron 
density away from half-filling and also include hopping between next nearest-neighbors. 
Second, they use an alternate definition for 
$\lambda = \alpha^2 N(E_{\text{F}})/M\Omega^2$, where $N(E_{\text{F}})$ is the 
density of states evaluated at the Fermi energy.
Nevertheless, the results in Fig.~\ref{fig:1}c-d reveal that the nonlinear 
corrections to the linear model are non-negligible at high temperature, even 
before the breakdown of Migdal's approximation becomes apparent.

%------------------------------------------------------------------------------%
\vskip 0.2cm
\noindent{\bf Pairing susceptibilities for large phonon frequency}.
Now we consider a counter comparison in the antiadiabatic regime with 
intermediate coupling by setting $ \lambda =0.4 $, $ \Omega = 4t $, 
$ N = 10 \times 10 $, and $ n = 0.55 $ (Fig.~\ref{fig:2}).
Away from half-filling, the pairing correlations grow more rapidly in part 
due to the larger $\Omega$, but also because of less competition with 
(incommensurate) CDW correlations (see Supplementary Note 3). 
Each of the curves in Fig.~\ref{fig:2} show that the system has strong pairing 
correlations, but they would yield very different estimates for $T_\text{c}$.  
The SCMA significantly overestimates $\chi^{\text{SC}}$, which is not 
surprising because Migdal's approximation is ill justified 
in this case (i.e., $ \lambda\frac{\Omega}{ E_{ \text{F} } } \sim 1 $). 
Interestingly, the nonlinear corrections become important at low temperature 
despite the presence of smaller lattice displacements (e.g. 
$ |\langle X_{i,l} \rangle|/A \approx 0.22 $).

\begin{figure}[t]
	\includegraphics[width=1.0\linewidth]{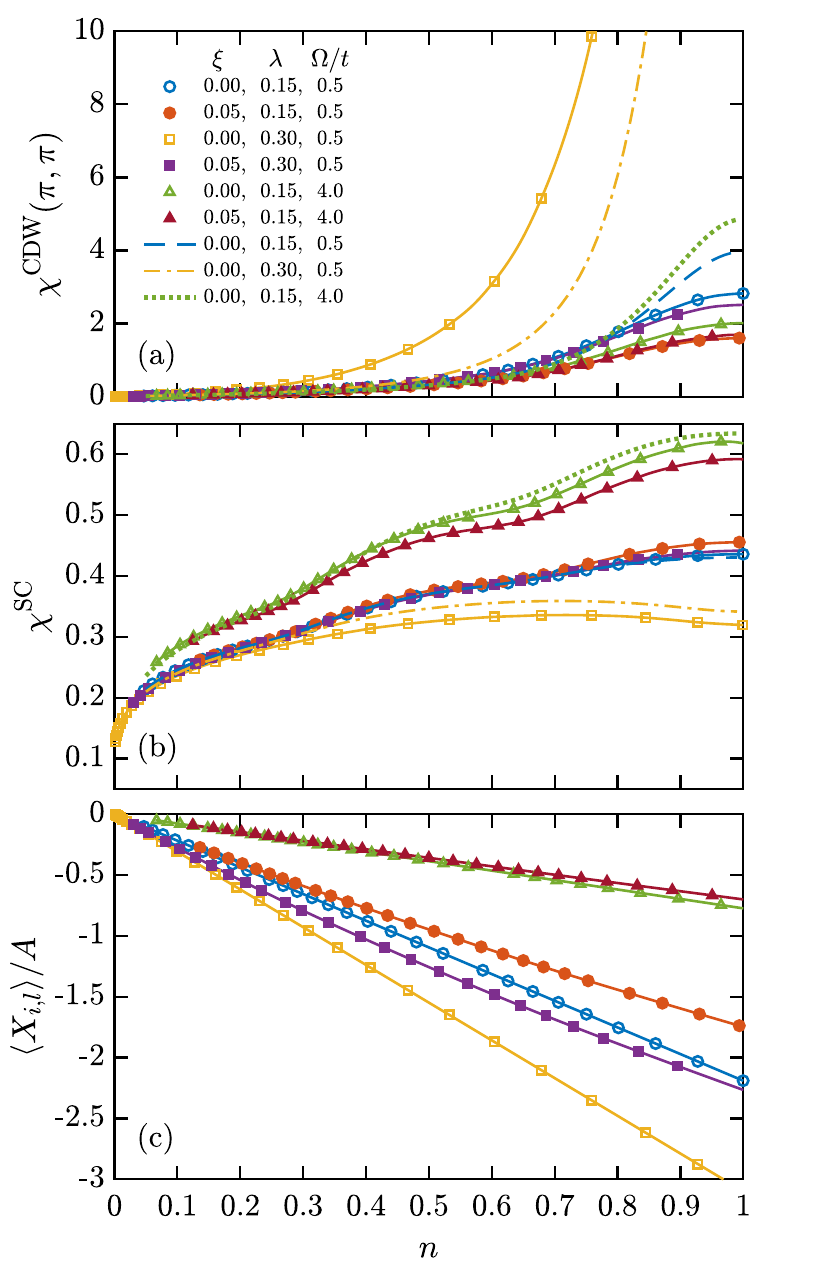}
	\caption{
		\textbf{Doping dependence of the correlations at fixed temperature 
		$\boldsymbol{T = 0.25t }$.}
		\textbf{(a)} The charge-density wave (CDW) susceptibility 
		$ \chi^{\text{CDW}}(\pi,\pi) $, 
		\textbf{(b)} superconducting (SC) pair-field susceptibility 
		$ \chi^{\text{SC}} $, and 
		\textbf{(c)} the average value of the phonon field 
		$ \langle X_{i,l} \rangle/A $ are shown for a lattice size of 
		$ N = 8\times 8 $. 
		Symbols connected by solid lines depict determinant quantum Monte Carlo 
		(DQMC) data where open (closed) symbols correspond to $ \xi = 0 $ 
		($ \xi = 0.05$). 
		Like symbol shapes between curves indicate the same pair of $\lambda$ 
		and $ \Omega $. 
		Dashed and dotted lines correspond to self consistent Migdal 
		approximation calculations for $ \xi = 0 $. 
		Error bars on the DQMC data were estimated using jackknife 
		resampling; however, all one-sigma error bars are smaller than the 
		symbol size and have been suppressed for clarity.
		Lines connecting DQMC data are used only to guide the eye.
	}
	\label{fig:3}
\end{figure}

%------------------------------------------------------------------------------%
\vskip 0.2cm
\noindent{\bf Comparison over doping}. 
Finally, Fig.~\ref{fig:3} shows results for three combinations of $ \lambda $ 
and $ \Omega/t $ over a wide range of electronic filling and at a fixed 
temperature $ T=0.25t $. 
The DQMC results for $ \xi = 0 $ ($\xi = 0.05$) are represented by open (closed) 
symbols in all three panels, while the SCMA results are shown as dashed or 
dotted lines in Fig.~\ref{fig:3}a-b. 
For reference, Fig.~\ref{fig:3}c shows the corresponding the average lattice 
displacement, obtained by averaging over {\it all} spacetime points 
$\langle X_{i,l} \rangle = \tfrac{1}{N^2L}\sum_{i,l}X_{i,l}$.
We caution that $\langle X_{i,l}\rangle$ provides a rough measure of the typical 
lattice displacements, and not a complete picture of the ionic subsystem.
We will return to this subtle issue later, when we discuss the displacement 
fluctuations.

%------------------------------------------------------------------------------%
{Case (1), $ \lambda = 0.15,\,\Omega = 0.5t $: }
These parameters are nearly identical to those used in Fig.~\ref{fig:1}. 
The deviations in $ \chi^{\text{CDW}}(\pi,\pi) $ for $ \xi = 0 $ and 
$\xi = 0.05$ (Fig.~\ref{fig:3}a) become apparent near $ n \geq 0.6 $ whereas 
the SCMA result starts to deviate from DQMC for $ n \geq 0.8 $.
At this temperature, the results for $ \chi^{\text{SC}} $ essentially agree
(Fig.~\ref{fig:3}b), but the nonlinear model yields a smaller average  
displacement $ \langle X_{i,l} \rangle/A $ (Fig.~\ref{fig:3}c).
These results further reinforce our prior observation that Migdal's 
approximation and the linear model can break down in different parameter regimes 
(in this case doping).

%------------------------------------------------------------------------------%
{Case (2), $ \lambda = 0.3,\,\Omega = 0.5t $: }  
Now we double $ \lambda $ while keeping the $\Omega$ fixed.
The increase in $ \lambda $ produces larger average displacements 
(Fig.~\ref{fig:3}c) and more pronounced nonlinear corrections. 
It also induces a stronger CDW (Fig.~\ref{fig:3}a) for 
the linear ($ \xi = 0 $) model. 
The SCMA qualitatively captures the CDW correlations of the linear model in 
panel (a), but underestimates their strength, which can be attributed to solely 
to the vertex corrections, consistent with the conclusions of Esterlis et 
al.~[\onlinecite{Esterlis2018}].
Due to the large CDW correlations, there is a 
suppression~\cite{Li2012,Marsiglio2019arXiv} in $ \chi^{\text{SC}} $ for 
$ \xi = 0 $, which is mostly captured by the SCMA (Fig.~\ref{fig:3}b). 
The introduction of the nonlinear interaction significantly reduces the CDW 
correlations and their competition with SC, which enhances $\chi^\mathrm{SC}$ at 
larger values of $n$.

%------------------------------------------------------------------------------%
{Case (3), $ \lambda = 0.15,\,\Omega = 4.0t $: } 
Now we look at the large phonon frequency results for DQMC (green and crimson 
triangles) and SCMA (green dotted line).
The larger $\Omega$ boosts pairing correlations (Fig.~\ref{fig:3}b) across the 
entire doping range and all of the $\chi^{\text{SC}}$'s are in fair agreement.
However, we know from Fig.~\ref{fig:2} that larger differences between each 
curve will emerge at lower temperatures.
In fact, the SCMA already overestimates $ \chi^{\text{CDW}}(\pi,\pi) $ near 
half-filling at this temperature, a feature we may attribute to 
antiadiabaticity.
The increased value of $\Omega$ means that the lattice vibrations are 
characterized by stiffer spring constants.
We obtain smaller average lattice displacements at all $n$ as a 
result~(Fig.~\ref{fig:3}c), which reduces the importance of the nonlinear 
interaction and produces better agreement between $ \xi = 0 $ and $ \xi = 0.05 $ 
DQMC results.

%------------------------------------------------------------------------------%
We should be careful in interpreting the results at lower filling in each of 
the cases above.
On the one hand, our examples suggest that corrections to the {\it e}-ph interaction 
are most important for describing the CDW phase transition near half-filling, 
which appears at higher temperatures. 
On the other hand, corrections could become important in the dilute carrier 
region at much lower temperatures. 
Nonetheless, our results suggest that the linear Holstein model is sensitive to 
nonlinear corrections over a large parameter space and that regions of this 
space overlap with regions where Migdal's approximation is not valid. 
But perhaps more importantly, there are regions where the linear approximation 
breaks down before Migdal's approximation does. 
%

%------------------------------------------------------------------------------%

\vskip 0.2cm
\noindent{\bf Average lattice displacement and its fluctuations}. 
In Fig.~\ref{fig:3}c we showed that the magnitude of the mean displacement 
$\langle X_{i,l} \rangle/A$ had an approximately linear dependence on the 
filling $ n $.
These displacements become larger when the dimensionless {\it e}-ph coupling 
$\lambda$ is increased or when the phonon energy $\Omega$ is decreased (or, 
equivalently, when the spring constants are softer). 
The behavior of $\langle X_{i,l}\rangle$ as a function of doping can be loosely 
understood by considering the atomic limit.
In this case, the effect of the linear {\it e}-ph-interaction is to shift the 
equilibrium position of the lattice to 
$X_{0} \approx -n\tfrac{\alpha_1}{M\Omega^2}$~\cite{JohnstonPRB2013}.
Indeed, the results shown in Fig.~\ref{fig:3}c for the linear model are well 
described by this function.
Based on this observation, one might then be tempted to try to eliminate the 
nonlinear interactions by defining new lattice operators 
$\hat{X}^\prime = \hat{X}-X_0=\hat{X}$ in hopes that the displacements of 
$\hat{X}^\prime$ remain small.
Unfortunately, this procedure is not viable for several reasons.

%------------------------------------------------------------------------------%
The first reason is that global shift of the equilibrium position will only be 
effective in the case of a uniform charge distribution.
This certainly will not be the case when the CDW correlations are significant.
For example, in the $(\pi,\pi)$ CDW phase, half of the sites are doubly 
occupied with an average displacement of $\approx 2X_0$ while the remaining 
sites are unoccupied with an average displacement of zero.
In this instance, $\langle X_{i,l}\rangle = X_0$, consistent with our results in 
Fig.~\ref{fig:3}c, but shifting the origin to $X=X_0$ will not eliminate the 
large lattice displacements at each site.

%------------------------------------------------------------------------------%
The second reason why redefining the origin will not work is that such 
transformations do not affect the displacement fluctuations, which are also 
significant for the linear Holstein model.
To show this, we examine the root-mean-square (rms) displacement of $X_{i,l}$ in 
our system with a linear {{\it e}-ph} coupling strength $\lambda$, which is defined as
\begin{equation}
    \sigma_{X}(\lambda) = \sqrt{\langle X_{i,l}^{2} \rangle 
    - \langle X_{i,l} \rangle^{2}}
\end{equation}
and is formally identical to a standard deviation. 
The value of $\sigma_{X}$ in the limit $ n \rightarrow 0$ 
approaches the (thermal) rms displacement for the free harmonic oscillator, 
which we denote as $\sigma_{X}(0)$ and is given by 
\begin{equation}
    \sigma_{X}(0) = \sqrt{\frac{1}{2\Omega}\left[ 2 n_{\text{B}}(\Omega) 
    + 1 \right]},
\end{equation}
where $n_{\text{B}}(\Omega) = [\mathrm{e}^{\beta \Omega} - 1]^{-1}$ is the Bose 
occupation function.  
%
%------------------------------------------------------------------------------%

\begin{figure}[t]
	\includegraphics[width=1.0\columnwidth]{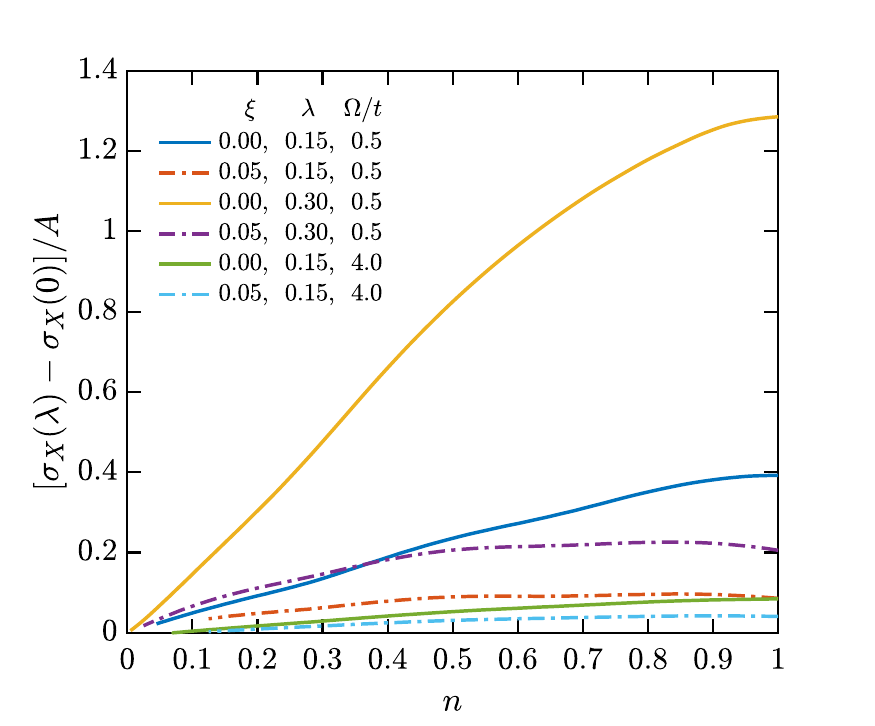}
	\caption{
		\textbf{Doping dependence of the root-mean-square (rms) lattice fluctuations.}
		Comparison of rms fluctuations beyond the (noninteracting) thermal 
		oscillator value, denoted $[\sigma_{X}(\lambda) - \sigma_{X}(0)]/A$ as a 
		function of filling for $T = t/4$ and 	$ N = 8 \times 8 $.
		Here, $\sigma_{X}(0)$ is the baseline contribution to the rms 
		fluctuations from the noninteracting thermal oscillator and 
		$\sigma_{X}(\lambda)$ is the fluctuation in the full interacting 
		problem. 
		Again, we find similar behavior to Fig.~\ref{fig:3}c only now we are 
		looking at the growth of oscillations beyond zero-point fluctuations 
		(which can also be large).
	}
	\label{fig:4}
\end{figure}

%------------------------------------------------------------------------------%

Fig.~\ref{fig:4} shows results for $[\sigma_X(\lambda)-\sigma_X(0)]/A$, as a 
function of filling for the same parameters used in Fig.~\ref{fig:3}.
(For reference, for $\Omega/t=0.5$ and $4.0$, we obtain 
$\sigma_X(0)/A = 1.146$ and $0.354$, respectively.) 
Here, we see that the fluctuations of the linear Holstein model are quite 
sensitive to the size of the dimensionless linear {\it e}-ph interaction $\lambda$.
Moreover, the magnitude of the fluctuations generally grow monotonically 
with filling until reaching a maximum at half-filling.
There, the largest fluctuations shown correspond to 
$\sigma_X(\lambda=0.3)/A\approx 2.43$, which is more than double the size of 
captured by $\sigma_X(0)/A$. Again, taking FeSe or a typical cuprate as 
references, these fluctuations correspond to $\sim 3.0\%$ and $\sim 6.3\%$ of 
the respective lattice constants. 
It is important to note that $\sigma_X(0)/A$ rises sharply for even smaller 
(and more realistic) model values of $\Omega/t\approx 0.02-0.1$. 
Such values, however, are typically inaccessible to DQMC due to prohibitively 
long autocorrelation times. 

The overall effect of the nonlinear coupling is to suppress the rms 
displacements fluctuations relative to the linear case, especially near 
half-filling. 
Only when $(\xi,\,\lambda,\,\Omega/t) = (0,\,0.15,\,4.0) $ and 
$(0.05,\,0.15,\,4.0)$ do we find close agreement between the linear and 
nonlinear models, and typical lattice displacements that are a small fraction 
of the lattice spacing. 

%----------------------------DFT Calc------------------------------------------%
\vskip 0.2cm
\noindent{\bf How big are nonlinear interactions in materials?}
Throughout this work we used $ \xi = 0.05 $, but how representative is this 
value for a realistic system?
To address this question, we considered the case of bulk FeSe, a quasi-2D 
material where the position of the Se atoms influence the on-site energies of 
the Fe 3$d$ orbitals~\cite{Gerber2017}, somewhat akin to the 2D Holstein model. 
To determine the strength of the linear and nonlinear {\it e}-ph coupling, we 
constructed a Wannier function basis from density functional theory (DFT) 
calculations to determine the on-site energy of the Fe 3d orbitals 
$ \epsilon_{3d}(z_{\text{Se}}) $ as a function of the Se atom's static 
displacement $ z_{\text{Se}} $ along the $c$-axis (see Supplementary Fig.~1). 
We then applied a polynomial fit of the form 
$ f(z_{\text{Se}}) = a_0 + a_{1}(z_{\text{Se}} - z_0) + a_{2}(z_{\text{Se}} 
- z_0)^{2} $ to the site energy $ \epsilon_{\text{diag}}(z_{\text{Se}}) $ for 
each orbital and computed $ \xi =A\frac{a_2}{a_1} $. 
Here, the oscillation amplitude $ A=\sqrt{ \hbar/2M\Omega }\approx 0.036$ \AA{}, 
adopting a selenium mass $ M = 1.31\times 10^{-25} $ kg, and the calculated 
phonon frequency $ \Omega = 2\pi\cdot5.02 $ THz of the A$_{1g}$ mode (See Model 
section). 
%

%------------------------------------------------------------------------------%
The results are summarized in Table \ref{tab1}, where $\xi$ ranges from 
$-0.1640$ to $0.0148$, with the strongest nonlinearity appearing for the 
$ d_{xy} $ orbital.      
We do not investigate $ \xi < 0 $ in our model calculations because others have 
shown that it leads to increased softening of the phonon dispersion and larger 
CDW correlations~\cite{Li2015b}.
Nevertheless, our results show $|\xi|\approx 0.05$ is certainly not out of the 
question for a real material. 

\begin{table}[]
\vspace{0.25cm}
\begin{tabular}{c  p{2.5cm}  p{2.5cm}}
\hhline{===}
%\hline
Orbital           & \hfil$a_2 / a_1$ [\AA{}$^{-1}$] & \hfil $ \xi $  \\ \hline
$ d_{xy} $        & \hfil -4.5936		& \hfil -0.1641 \\
$ d_{xz}/d_{yz} $ & \hfil -0.4804		& \hfil -0.0172 \\
$ d_{z^{2}} $     & \hfil 0.0159		& \hfil 0.0006  \\
$ d_{x^2 - y^2} $ & \hfil 0.4155		& \hfil 0.0148  \\ \hhline{===}
\end{tabular}
\caption{\textbf{Estimation of nonlinear {\it e}-ph coupling ratio $ \xi $ in FeSe.}
Results from fits of the on-site energy 
$ \epsilon_{\text{d}}(z_{\text{Se}}) $ for each $d$-orbital of Fe as a function 
of the height of the Se atom $z_{\text{Se}}$ measured with respect to the 
Fe-plane. 
Fitting $ \epsilon_{\text{d}}(z_{\text{Se}}) $ with a simple polynomial of the 
form $ f(z_{\text{Se}}) = a_0 + a_{1}(z_{\text{Se}} - z_0) + a_{2}(z_{\text{Se}} 
- z_0)^{2} $, we estimate the nonlinear coupling ratio $ \xi $ from the fitting 
parameters $ a_{2}/a_{1} $.  
}
\label{tab1}
\end{table}

%------------------------------------------------------------------------------%
%----------------------------- Conclusion -------------------------------------%
%------------------------------------------------------------------------------% 
\section*{Discussion}  
We have demonstrated that the linear approximation to the {\it e}-ph interaction in 
the Holstein model breaks down in commonly studied parameter regimes.
Importantly, this breakdown regime overlaps with ones where Migdal's 
approximation captures the DQMC result, even if only qualitatively.
This observation indicates that nonlinear corrections to the underlying linear 
lattice model may be important even when vertex corrections are not.
We also studied the example of bulk FeSe from first principles and found that 
nonlinear {\it e}-ph interactions in a real materials can be quite significant and on 
par with, or even larger than our model choice of $ |\xi| =0.05 $. 

%------------------------------------------------------------------------------%
It is natural to wonder which parameter regimes might be best for ensuring 
lattice displacements remain small enough justify the use of a linear 
interaction. 
We have found that tuning $ \lambda $ to smaller values suppresses the lattice 
displacements and their fluctuations, but also pushes the growth of correlations 
to lower temperatures, making computations more expensive.
(Some groups~\cite{Hohenadler2019} have recently managed to access such 
temperatures in QMC, however.) 
Alternatively, one could also shrink the displacements by choosing antiadiabatic 
parameters (i.e., $ \Omega > E_{\text{F}} $). 
But even for a strongly antiadiabatic choice of 
$ (\lambda,\, \Omega/t,\, n) = (0.4,\,4.0 ,0.55) $, nonlinear corrections to the 
{\it e}-ph interaction produced considerable differences in the resulting 
temperature dependence of the superconducting susceptibility. 
Unfortunately, focusing on smaller phonon energies, which are relevant for real 
materials, will also produce larger lattice displacements and fluctuations that 
are inconsistent with a linear interaction.
While our results are not comprehensive across the entire parameter space of 
the Holstein model, we are forced to conclude that they do call large portions 
of this space into question. 
For instance, our results imply that combinations of $ \lambda \gtrsim 10^{-1} $ 
and $ \Omega \lesssim 4t $ yield sizable displacements and displacement 
fluctuations, which would necessitate additional nonlinear interactions and/or 
anharmonic lattice potentials~\cite{FreericksPRL1997}.  
Our results indicate a clear and present need for more work extending beyond the 
simplest effective models, especially when one is trying to describe the physics 
of a real system. 

%------------------------------------------------------------------------------%
The Holstein model and Migdal's approximation have long served as cornerstones 
in the study of electron-phonon interactions.
Their relative simplicity has helped shape our intuition about superconductivity, 
its competition with charge order, and polaron formation, and studying the 
Holstein model can address the essential physics of these processes.
% %
While it is clear that these models are built on the assumption of small lattice 
displacements, it is not always clear how large these displacements will be in 
practice or whether additional nonlinear interactions will modify the physics of 
the model.
One must, therefore, be careful when extrapolating results from effective models 
to real materials when they are driven outside their range of validity. 
For example, we have shown that the Holstein model can produce displacements 
that begin to approach the Lindemann criteria for melting (particularly as 
$\Omega$ is reduced), but the model cannot describe such a transition. 
Instead, it over predicts various tendencies towards ordered phases in these 
cases. 
Similarly, it is unclear how one should map critical $\lambda$ values 
derived for the breakdown of Migdal's approximation onto real materials.

%------------------------------------------------------------------------------%
%------------------------------- Methods --------------------------------------%
%------------------------------------------------------------------------------% 
\section*{Methods}
\noindent{\bf Determinant quantum Monte Carlo}.  
DQMC is an auxiliary field, imaginary-time technique that computes expectation 
values within the grand canonical ensemble. 
It is inherently nonperturbative and includes all Feynman diagrams.
While QMC simulations of the (non)linear Holstein model face long 
autocorrelation times~\cite{Hohenadler}, they are free of a sign problem. 
We refer the reader to Supplementary Note 2 for more details on our DQMC 
implementation and more generally to White {\it et al}.~\cite{WhitePRB1989}, 
Scalettar {\it et al}.~\cite{Scalettar1989}, and Johnston {\it et al}.~\cite{JohnstonPRB2013}.

\vskip 0.5cm
\noindent{\bf The self-consistent Migdal approximation}. 
The self-consistent Migdal approximation (SCMA) is a diagrammatic approach that 
neglects higher-order corrections to the {\it e}-ph interaction vertex. 
Here, we use a recently developed SCMA code that treats both the electron and 
phonon self-energies on an equal footing and captures the competition between 
the CDW and SC instabilities~\cite{DeePRB2019}. 
%

%------------------------------------------------------------------------------%
\vskip 0.5cm
\noindent{\bf Susceptibilities}.  
The effects of nonlinear {\it e}-ph coupling or the omission of vertex corrections 
will manifest uniquely in different observables. 
Here, we focus on two-particle correlation functions. 

%------------------------------------------------------------------------------%
The CDW correlation at momentum $ {\bf q} $ is measured by the charge  
susceptibility
\begin{equation}\label{eq:CDW}
\chi^{\text{CDW}}({\bf q}) = \frac{1}{N} \int_{0}^{\beta} \mathrm{d} \tau 
			\,\langle \hat{\rho}^{{\phantom{\dagger}}}_{{\bf q}}(\tau) \hat{\rho}^{\dagger}_{{\bf q}}(0) \rangle_{\text{c}},
\end{equation}
where $ \hat{\rho}^{{\phantom{\dagger}}}_{{\bf q}}(\tau) \equiv \sum_{i,\sigma} 
\mathrm{e}^{-\mathrm{i}{\bf q}\cdot{\bf R}_i}\,\hat{n}_{i,\sigma}(\tau) $ and 
$\langle \hat{A}\hat{B} \rangle_{\text{c}} = \langle \hat{A}\hat{B} \rangle - 
\langle \hat{A} \rangle \langle \hat{B} \rangle$ denotes a connected
correlation function.
Near half-filling, the Fermi surface is well nested and the charge 
susceptibility has a single commensurate peak at 
$ {\bf q}_{\text{max}}=(\pi,\,\pi)  $.  
Moving away from half-filling removes the nesting condition and eventually 
redistributes the weight of the single peak in $ \chi^{\text{CDW}}({\bf q})  $ into 
four incommensurate peaks (see Supplementary Fig.~3).
%

%------------------------------------------------------------------------------%
The {\it e}-ph coupling is also responsible for spin-singlet $ s $-wave pairing, 
resulting in superconducting correlations measured by the pair-field 
susceptibility 
\begin{equation}
	\chi^{\text{SC}} = \frac{1}{N}\int_{0}^{\beta}\mathrm{d}\tau 
	\,\langle \hat{\Delta}(\tau) \hat{\Delta}^{\dagger}(0) \rangle,
\end{equation} 
where 
$ \hat{\Delta}(\tau)=\sum_{i}\hat{c}^{{\phantom{\dagger}}}_{i,\uparrow}(\tau)\hat{c}^{{\phantom{\dagger}}}_{i,\downarrow}(\tau)$.

\vspace{0.2cm}

\noindent
\textbf{Data availability}: 
Data are available upon request.

\vspace{0.2cm}

\noindent
\textbf{Code availability}: 
The SMCA code is available at \url{https://github.com/johnstonResearchGroup/Migdal}.
The DQMC code is available upon request.

\vspace{0.2cm}

\noindent
\textbf{Acknowledgments}: 
We thank M. Berciu and D.~J. Scalapino for providing early feedback on the 
manuscript.
We also thank B. Cohen-Stead. and R.~T. Scalettar for valuable discussions 
regarding the length scales discussed in this work.
This work is supported by the Scientific Discovery through
Advanced Computing (SciDAC) program funded by the U.S. Department of Energy, 
Office of Science, Advanced Scientific Computing Research and Basic Energy 
Sciences, Division of Materials Sciences and Engineering.
J.~C. recognizes the support of the DOE Computational Science Graduate 
Fellowship (CSGF) under grant DE-FG02-97ER25308.
This research also used resources of the Oak Ridge Leadership Computing 
Facility, which is a DOE Office of Science User Facility supported under 
Contract DE-AC05-00OR22725.

\vspace{0.2cm}
\noindent
\textbf{Author contributions}: 
P.~D. performed DQMC and SCMA calculations and the associated analysis. 
K.~K also performed DQMC calculations. 
J.~C. performed DFT calculations and the associated analysis. 
P.~D., J.~C., and S.~J. wrote the manuscript. 
S.~J. conceived of the project and supervised the work.

\vspace{0.2cm}
\noindent
\textbf{Corresponding author}: Requests for materials should be directed to S.~J.~(email: sjohn145@utk.edu).  

\vspace{0.2cm}
\noindent
\textbf{Competing interests}: The authors declare no competing interests.


\begin{thebibliography}{99}
\bibitem{Giustino2017}
Feliciano, G., Electron-phonon interactions from first principles. {\it Rev. Mod. Phys.} {\bf 89}, 015003 (2017). \url{https://link.aps.org/doi/10.1103/RevModPhys.89.015003}

\bibitem{Migdal1958}
Migdal, A. B., Interaction between electrons and lattice vibrations in a normal metal, {\it Sov. Phys. JETP}, {\bf  7}, 996--1001 (1958).

\bibitem{Eliashberg1960}
Eliashberg, G. M., {Interactions between electrons and lattice vibrations in a superconductor}. {\it Sov. Phys. JETP} {\bf 11}, 696--702 (1960).

\bibitem{Eliashberg1961}
Eliashberg, G. M., Temperature Green's Function For Electrons In a Superconductor. {\it Sov. Phys. JETP} {\bf 12}, 1000--1002 (1961).
 
\bibitem{Freericks1997}
Freericks, J. K., Nicol, E. J., Liu, A. Y., and Quong, A. A., {Vertex-corrected tunneling inversion in electron-phonon mediated superconductors: Pb}. {\it Phys. Rev. B}, {\bf 55}, 11651--11658 (1997). \url{https://link.aps.org/doi/10.1103/PhysRevB.55.11651}

\bibitem{Alexandrov2001}
Alexandrov, A. S., {Breakdown of the Migdal-Eliashberg theory in the strong-coupling adiabatic regime}.  {\it  Europhys. Lett.} 
{\bf 56}, 92--98 (2001). \url{https://iopscience.iop.org/article/10.1209/epl/i2001-00492-x}

\bibitem{Hague2003}
Hague, J. P.,  {Electron and phonon dispersions of the two-dimensional Holstein model: effects of vertex and non-local corrections}.
{\it Journal of Physics: Condensed Matter} {\bf 15}, 2535--2550
(2003).  \url{https://iopscience.iop.org/article/10.1088/0953-8984/15/17/309}

\bibitem{Bauer2011}
Bauer, J., Han, J. E., and Gunnarsson, O., 
Quantitative reliability study of the Migdal-Eliashberg theory for strong electron-phonon coupling in superconductors. 
{\it Phys. Rev. B}, {\bf 84}, 184531 (2011). 
\url{https://link.aps.org/doi/10.1103/PhysRevB.84.184531}

\bibitem{Esterlis2018}
Esterlis, I., Nosarzewski, B., Huang, E. W., Moritz, B., Devereaux, T. P., Scalapino, D. J., and Kivelson, S. A., Breakdown of the Migdal-Eliashberg theory: A determinant quantum Monte Carlo study. {\it Phys. Rev. B} {\bf 97}, 140501(R) (2018). 
\url{https://link.aps.org/doi/10.1103/PhysRevB.97.140501}

\bibitem{Liu2019}
Liu, G.-Z.-Zhu, Yang, Z.-K., Pan, X.-Y., and
Wang, J.-R., Towards exact solutions of electron-phonon interaction in metals. (2019). Preprint available at arXiv:1911.05528. 
\url{https://arxiv.org/abs/1911.05528} 

\bibitem{Schrodi2019}
Schrodi, F., Oppeneer, P. M., and Aperis, A.,
Full-bandwidth Eliashberg theory of superconductivity beyond Migdal's approximation. (2019). Preprint available at arXiv:1911.12872.
\url{https://arxiv.org/abs/1911.12872}. 

\bibitem{Gastiasoro2019}
Gastiasoro, Maria N., Chubukov, Andrey V., and Fernandes, Rafael M., Phonon-mediated superconductivity in low carrier-density systems. {\it Phys. Rev. B} {\bf 99}, 094524 (2019). \url{https://link.aps.org/doi/10.1103/PhysRevB.99.094524}

\bibitem{Holstein1959}
Holstein, T., Studies of polaron motion: Part I. The molecular-crystal model. {\it Annals of Physics} {\bf 8}, 325 -- 342 (1959). 
\url{http://www.sciencedirect.com/science/article/pii/0003491659900028}

\bibitem{Frohlich1954}
{Fr{\"o}hlich}, H., Electrons in lattice fields. {\it Advances in Physics} {\bf 3}, 325--361 (1954). 

\bibitem{Scalettar1989}
Scalettar, R. T., Bickers, N. E., and Scalapino, D. J., Competition of pairing and Peierls--charge-density-wave correlations in a two-dimensional electron-phonon model. {\it Phys. Rev. B} {\bf 40}, 197--200 (1989). \url{https://link.aps.org/doi/10.1103/PhysRevB.40.197}

\bibitem{Marsiglio1990}
Marsiglio, F., Pairing and charge-density-wave correlations in the Holstein model at half-filling. {\it Phys. Rev. B} {\bf 42}, 2416--2424 (1990). \url{https://link.aps.org/doi/10.1103/PhysRevB.42.2416}

\bibitem{Levine1990}
{Levine, G., and Su, W. P.}, {Pairing of charge carriers in the two-dimensional molecular crystal model}. {\it Phys. Rev. B} {\bf 42},
{4143--4149} (1990). \url{https://link.aps.org/doi/10.1103/PhysRevB.42.4143}

\bibitem{Levine1991}
{Levine, G., and Su, W. P.}, {Finite-cluster study of superconductivity in the two-dimensional molecular-crystal model}. {\it Phys. Rev. B}, {\bf 43}, {10413--10421} (1991). \url{https://link.aps.org/doi/10.1103/PhysRevB.43.10413}

\bibitem{Noack1991}
{Noack, R. M., Scalapino, D. J., and Scalettar, R. T.}, {Charge-density-wave and pairing susceptibilities in a two-dimensional
	electron-phonon model}. {\it Phys. Rev. Lett.}, {\bf 66}, {778--781} (1991). \url{https://journals.aps.org/prl/abstract/10.1103/PhysRevLett.66.778}

\bibitem{Vekic1992}
{Veki\'{c}, M., Noack, R. M., and White, S. R.}, {Charge-density waves versus superconductivity in the Holstein model with next-nearest-neighbor hopping}. {\it Phys. Rev. B} {\bf 46}, 271--278 (1992). \url{https://journals.aps.org/prb/abstract/10.1103/PhysRevB.46.271}

\bibitem{Vekic1993}
{Veki\ifmmode \acute{c}\else \'{c}\fi{}, M., and White, S. R.}, {Gap formation in the density of states for the Holstein model}. {\it Phys. Rev. B}, {\bf 48}, 7643--7650 (1993). \url{https://link.aps.org/doi/10.1103/PhysRevB.48.7643}

\bibitem{Niyaz1993}
{Niyaz, P., and Gubernatis, J. E., Scalettar, R. T., and Fong, C. Y.}, {Charge-density-wave-gap formation in the two-dimensional Holstein model at half-filling}. {\it Phys. Rev. B} {\bf 48}, 
{16011--16022} (1993).  \url{https://journals.aps.org/prb/abstract/10.1103/PhysRevB.48.16011}

\bibitem{Freericks1995PRL}
{Freericks, J. K., and Jarrell, Mark}, {Competition between Electron-Phonon Attraction and Weak Coulomb Repulsion}. {\it Phys. Rev. Lett.} {\bf 75}, {2570--2573} (1995).  \url{https://link.aps.org/doi/10.1103/PhysRevLett.75.2570}

\bibitem{FreericksPRL1997}
{Freericks, J. K., Jarrell, M., and Mahan, G. D.}, The Anharmonic Electron-Phonon Problem [Phys. Rev. Lett. 77, 4588 (1996)]. {\it Phys. Rev. Lett.} {\bf 79},  {1783--1783} (1997).  \url{https://link.aps.org/doi/10.1103/PhysRevLett.79.1783}

\bibitem{Zheng1997}
{Zheng, H., and Zhu, S. Y.}, {Charge-density-wave and superconducting states in the Holstein model on a square lattice}. {\it Phys. Rev. B},
  volume = {\bf 55}, {3803--3815} (1997). \url{https://link.aps.org/doi/10.1103/PhysRevB.55.3803}

\bibitem{Goodvin2006}
{Goodvin, G. L., and Berciu, Mona and Sawatzky, George A.}, {Green's function of the Holstein polaron}. {\it Phys. Rev. B} {\bf 74},  {245104} (2006). \url{https://link.aps.org/doi/10.1103/PhysRevB.74.245104}

\bibitem{Chen2018}
{Chen, C., Xu, Xiao Y., Liu, J., and Batrouni, G., Scalettar, R. T., and Meng, Z. Y.}, {Symmetry-enforced self-learning Monte Carlo method applied to the Holstein model}. {\it Phys. Rev. B}, {\bf 98}, {041102(R)} (2018). \url{https://link.aps.org/doi/10.1103/PhysRevB.98.041102}

\bibitem{Li2019}
{Li, S., Dee, P, M., Khatami, E., and Johnston, S.}, {Accelerating lattice quantum Monte Carlo simulations using artificial neural networks: Application to the Holstein model}. {\it Phys. Rev. B}, {\bf 100}, 020302(R) (2019). \url{https://link.aps.org/doi/10.1103/PhysRevB.100.020302}

\bibitem{Hohenadler2019}
{Hohenadler, M., and Batrouni, G. G.}, {Dominant charge density wave correlations in the Holstein model on the half-filled square lattice}. {\it Phys. Rev. B}, {\bf 100}, {165114} (2019). \url{https://link.aps.org/doi/10.1103/PhysRevB.100.165114}

\bibitem{Devreese1996}
{Devreese, J. T.}, {Polarons}. {\it Encyclopedia of Applied Physics} {\bf 14}, 383 (1996).  

\bibitem{Adolphs2013}
{Adolphs, C. P. J.}, and {Berciu, M.}, {Going beyond the linear approximation in describing electron-phonon coupling: Relevance for the Holstein model}, {\it {EPL} (Europhysics Letters)} {\bf 102}, 47003 (2013). \url{https://doi.org/10.1209/0295-5075/102/47003}   

\bibitem{Li2015a}
{Li, S. and Johnston, S.},
{The effects of non-linear electron-phonon interactions on superconductivity and charge-density-wave correlations}. 
{\it {EPL} (Europhysics Letters)} {\bf 105}, 27007 (2015). 
\url{http://dx.doi.org/10.1209/0295-5075/109/27007}

\bibitem{Li2015b}
{Li, S., Nowadnick, E. A., and Johnston, S.},  {Quasiparticle properties of the nonlinear Holstein model at finite doping and temperature}. {\it Phys. Rev. B}, {\bf 92}, 064301 (2015). 
\url{https://link.aps.org/doi/10.1103/PhysRevB.92.064301}

\bibitem{WhitePRB1989}
{White, S. R., Scalapino, D. J., Sugar, R. L., Loh, E. Y., Gubernatis, J. E., and Scalettar, R. T.}, {Numerical study of the two-dimensional Hubbard model}. {\it Phys. Rev. B} {\bf 40}, 
506--516 (1989)\url{https://link.aps.org/doi/10.1103/PhysRevB.40.506}

\bibitem{Li2012}
Li, Z., Marsiglio, F., The Polaron-Like Nature of an Electron Coupled to Phonons. {\it Journal of Superconductivity and Novel Magnetism} {\bf 25}, 1313--1317(2012). 
\url{https://doi.org/10.1007/s10948-012-1601-6}

\bibitem{Marsiglio2019arXiv}
Marsiglio, F., Eliashberg {Theory}: a short review. (2019).
Preprint available at arXiv:1911.05065.

\bibitem{JohnstonPRB2013}
Johnston, S., Nowadnick, E. A., Kung, Y. F., Moritz, B., Scalettar, R. T., and Devereaux, T. P., Determinant quantum Monte Carlo study of the two-dimensional single-band Hubbard-Holstein model. {\it Phys. Rev. B} {\bf 87}, 235133 (2013). 
\url{https://link.aps.org/doi/10.1103/PhysRevB.87.235133}

\bibitem{Gerber2017}
Gerber, S. {\it et al.}, Femtosecond electron-phonon lock-in by photoemission and x-ray free-electron laser. {\it  Science} {\bf 357},
{71--75} (2017). \url{https://science.sciencemag.org/content/357/6346/71}

\bibitem{Hohenadler}
Hohenadler, M., and Lang, T. C., Autocorrelations in Quantum Monte Carlo Simulations of Electron-Phonon Models. In {\it Computational Many-Particle Physics}, pages 357--366 (2008). 

\bibitem{DeePRB2019}
{Dee, P. M., Nakatsukasa, K., Wang, Y., and Johnston, S.}, {Temperature-filling phase diagram of the two-dimensional Holstein model in the thermodynamic limit by self-consistent Migdal approximation}. {\it Phys. Rev. B} {\bf 99}, 024514 (2019). 
\url{https://journals.aps.org/prb/abstract/10.1103/PhysRevB.99.024514}

\end{thebibliography}
\end{document}